\documentclass[aps,pre,twocolumn,superscriptaddress]{revtex4}  
\usepackage{graphicx,epsfig}  
\usepackage{dcolumn}   
\usepackage{bm}        
\usepackage{amssymb}   
\usepackage{color}
\usepackage{mathtools}
\usepackage{esvect}
\begin{document}

\title{Chaotic  dynamics of small sized charged Yukawa Dust Clusters}

\author{Priya Deshwal}
\email {priyadeshwal25@gmail.com} 	
\affiliation{Department of Physics, Indian Institute of Technology Delhi, Hauz Khas, New Delhi 110016, India}	
\author{Mamta Yadav}
\affiliation{Department of Physics, Indian Institute of Technology Delhi, Hauz Khas, New Delhi 110016, India}

\author{Chaitanya Prasad}
\affiliation{Computer Science Department, Ashoka University, Sonepat -131029, Haryana, India}
\author{Shantam Sridev}
\affiliation{Electronics and Computer Science Department, University of Southampton, Southampton SO17 1BJ, UK}
\author{Yash Ahuja}
\affiliation{Mechanical Engineering Department, Georgia Institute of Technology, Atlanta - 30313, Georgia, USA}
\author{Srimanta Maity}
\affiliation{Department of Physics, Indian Institute of Technology Delhi, Hauz Khas, New Delhi 110016, India}
\author{Amita Das}
\email {amita@iitd.ac.in}
\affiliation{Department of Physics, Indian Institute of Technology Delhi, Hauz Khas, New Delhi 110016, India}

\begin{abstract}
In a recent work, \cite{maity2020dynamical} the equilibrium of a cluster of charged dust particles  mutually interacting with 
screened Coulomb force and radially confined by an externally applied electric field in a 2-D configuration was studied. 
It was shown that the particles arranged themselves on discrete radial rings forming a lattice structure. 
In some cases with the specific number of particles, no static equilibrium was observed; instead, angular rotation of particles positioned at various rings was observed.  
In a two-ringed structure, it was shown that the direction of rotation was opposite. The direction of rotation was also observed to 
change apparently at random time intervals. 
A detailed characterization of the dynamics of small-sized Yukawa clusters has been carried out in the present work.   In particular,  it has been shown that the 
dynamical time reversal of angular rotation exhibits chaotic behavior.

\end{abstract}

\maketitle

 \section{\it Introduction}
Plasma medium being intrinsically nonlinear has attracted considerable interest in the study of chaotic dynamics associated with it \cite{fan1992observations,mitra2014order,shaw2019continuous}. These studies have mostly explored the chaotic behavior associated with macroscopic signals such as the current-voltage characteristics etc. Some recent studies \cite{sheridan2005chaos, sheridan2010transition} have, however, also shown the presence of chaos in the dynamics of the small charged cluster of particles immersed in a plasma medium. For instance, in a paper by T.E. Sheridan \cite{sheridan2005chaos}, it was shown that a three-particle system exhibits chaotic dynamics in the presence of a low-frequency modulation of the underlying background plasma density.  

The charged clusters are essentially dust particles immersed in plasma and form the basis for complex dusty plasma tabletop experiments \cite{juan1998observation}. The lighter electron species, in this case, gets attached to these micro-particles rendering them negatively charged.  The charge on the micro-particles can often be quite large, of the order of $10^4$ electronic charges. The experiments involving dusty plasma are fairly simple, and they can be easily pushed into the strongly coupled regime \cite{murillo2004strongly, chu1994direct, thomas1994plasma, hayashi1994observation} even at room temperature and normal densities. The dust charge gets shielded in the plasma environment, and as a result, the inter-dust interaction is typically described by the screened Coulomb potential \cite{shukla2015introduction}.
The experiments involving such configurations of dust particles are easy to perform, and the trajectory of individual dust particles can also be tracked with sufficient detail and with very simple diagnostics. Such a system offers an ideal test-bed for studying crystallization process \cite{chu1994direct, thomas1994plasma, hayashi1994observation}, single-particle dynamics \cite{juan1998anomalous, teng2009wave, maity2018interplay}, phase transitions \cite{melzer1996experimental, schweigert1998plasma, maity2019molecular}, etc.  
The system thus provides a very convenient example for studying the dynamical behavior of a small cluster of interacting particles placed in any externally applied field.

In a recently published work from our group \cite{maity2020dynamical} the 2-D equilibrium study of charged dust cluster immersed in plasma under a radially confining static force has been studied with the help of Molecular Dynamics simulations. The screening of the charged dust particles by the background plasma was accounted for by considering a screened Yukawa interaction amidst dust particles. The dust particles were shown to arrange themselves in interesting patterns at various radial rings. The system was observed to relax towards static configuration with particles placed in radial rings in a definite pattern. In some cases, depending on the number of particles, it was noticed that there was no static configuration possible. In such cases, the dust particles exhibited azimuthal rotation, and the direction of rotation kept changing with time in a seemingly random fashion. We provide here an understanding of the formation of these  structures 
on the basis of minimization of the total potential energy of the system. 
Furthermore, we explore the dynamical state exhibited by these small clusters of dust particles in detail and show that the azimuthal dynamics of the particles are essentially chaotic in time.

The paper has been organized as follows. In section II, we discuss the simulation details. Section III contains the details of possible configurations to which the system is observed to relax. The understanding and relaxation towards a dynamical state for some cases for which no static equilibrium exists has been provided in this section. In section IV, we study the dynamics in detail for some specific clusters and show that the azimuthal dynamics of the particles exhibit chaotic behavior. Section V contains the summary and conclusion.

 \label{intro}


\section{\it MD Simulation Details}
\label{mdsim}

We simulate of 2-D dust Yukawa cluster with a finite number of charged particles using classical Molecular Dynamics code, LAMMPS \cite{plimpton1995fast}. Each simulation is started with random phase-space distribution of dust grains inside the simulation box of normalized length $L = L_x = L_y = 12.79\lambda_0$ in $x$ and $y$ directions respectively. Here, $\lambda_0 = 2.2854\times 10^{-3}m$ is choose to normalized the length scales. We consider identically charged dust grains immersed in background plasma. The mass $(M)$ and charge $(Q)$ of dust species are taken to be $6.99\times 10^{-13}kg$ and $11940e$ respectively where $e$ is the charge of an electron.  
These values correspond to a typical experiment of dusty plasma \cite{nosenko2004shear}. All negatively charged dust particles repel each other via screened coulomb potential, $U(r) = (Q/4\pi\epsilon_0r)\exp{(-r/\lambda_D)}$ and are confined  in the $x-y$ plane by a parabolic potential which is provided by an externally applied electric field of the form $\mathbf{E} = K(x-L/2)\hat{x}+ K(y-L/2) \hat{y}$ . Here, $\lambda_{D}$ and $K$ are the typical Debye screening length and strength of confining potential respectively . We define normalized screening parameter defined as $(\kappa = \lambda_0/\lambda_D =1$ ) which represents the strength of pair interaction.  
  The value of $K$, which defines the strength of the electric field, has been chosen to be $2500$. Dynamics of dust particles are tracked by choosing time step $0.001 \omega_{0}^{-1}$ for simulations where $\omega_{0}=(QK/M)^{1/2} = 2.616 s^{-1}$. 
  The net force acting on any (say $i^{th}$) particle is sum of forces due to all other particles and external confinement force as given by the expression below
\begin{equation}
\mathbf{F}_{i} = -Q\sum_{j=1}^{N_p} \nabla U(\mathbf{r}_i, \mathbf{r}_j) + Q(\mathbf{E}_{x} + \mathbf{E}_{y}),
\label{total_frc}
\end{equation}
 A Nose-Hoover thermostat \cite{nose1984molecular,hoover1985canonical} is used to keep the system at the desired temperature, and phase space coordinates are generated from canonical ensemble using a thermostat. In addition, a chain of thermostats has been coupled to a particle thermostat \cite{plimpton1995fast,hoover1985canonical}. A thermal equilibrium state is obtained using a Nose-Hoover thermostat at particle kinetic temperature 416.95 K, which corresponds to Coulomb coupling parameter $(\Gamma)$ 2500. The typical particle  velocity $(v_{th})$ 
 corresponding to this temperature is $9.07\times 10^{-5} ms^{-1}$.

\section{\it Configuration of small clusters}
We studied the relaxation of a specific number of charge particles distribution randomly placed in the box. The configuration as expected tries to relax to an equilibrium state in which the energy is minimum. 
A single particle would thus always reside at the center of simulation box where the external potential energy is minimum. 
As the number of particles are  increased in the cluster  the interaction potential amidst the particles also becomes relevant.  
The repulsive Yukawa potential of the particles  tries to place  particles as far apart as possible whereas the  external electric field confines all of them close to the central  region of the box.  As a result of this the  particles arrange themselves in patterns as illustrated in  Table \ref{table}.  
The table shows   different arrangements to which the particle relax  as one changes the   number of particles. 
It can be observed from the table that when the particle number lies between $2$ to $5$ they are arranged in a single ring 
at equi-angular spacing. As the number of particles is increased,  we first get  a configuration in which a single particle is placed at the center and others are placed around  a ring at a particular radius. This continues till the total number of particle is $8$. For particle number $9$ and beyond the structures become more complicated. The inner shell now comprises of $2$ to $5$ particles and the rest of the particles are arranged at a larger radius. It is in these configurations  
that one observes that in most cases the  structure never relaxes to a stationary pattern, instead particles are observed to exhibit rotation. 

 \begin{figure}[hbt!]
   \includegraphics[height = 2.5cm,width = 8.0cm]{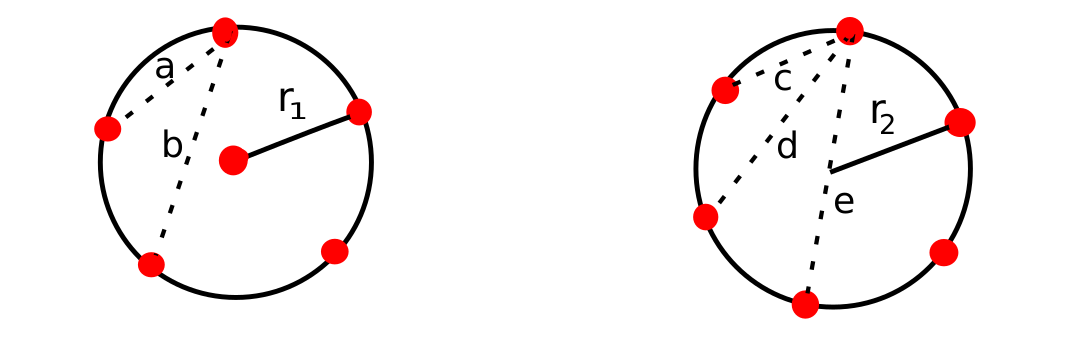}
   \caption{ (1,5) and (0,6) configurations of cluster having six particles. }
  \label{confi}
\end{figure}

 \begin{figure}[hbt!]
 \includegraphics[height = 7.5cm,width = 9.0cm]{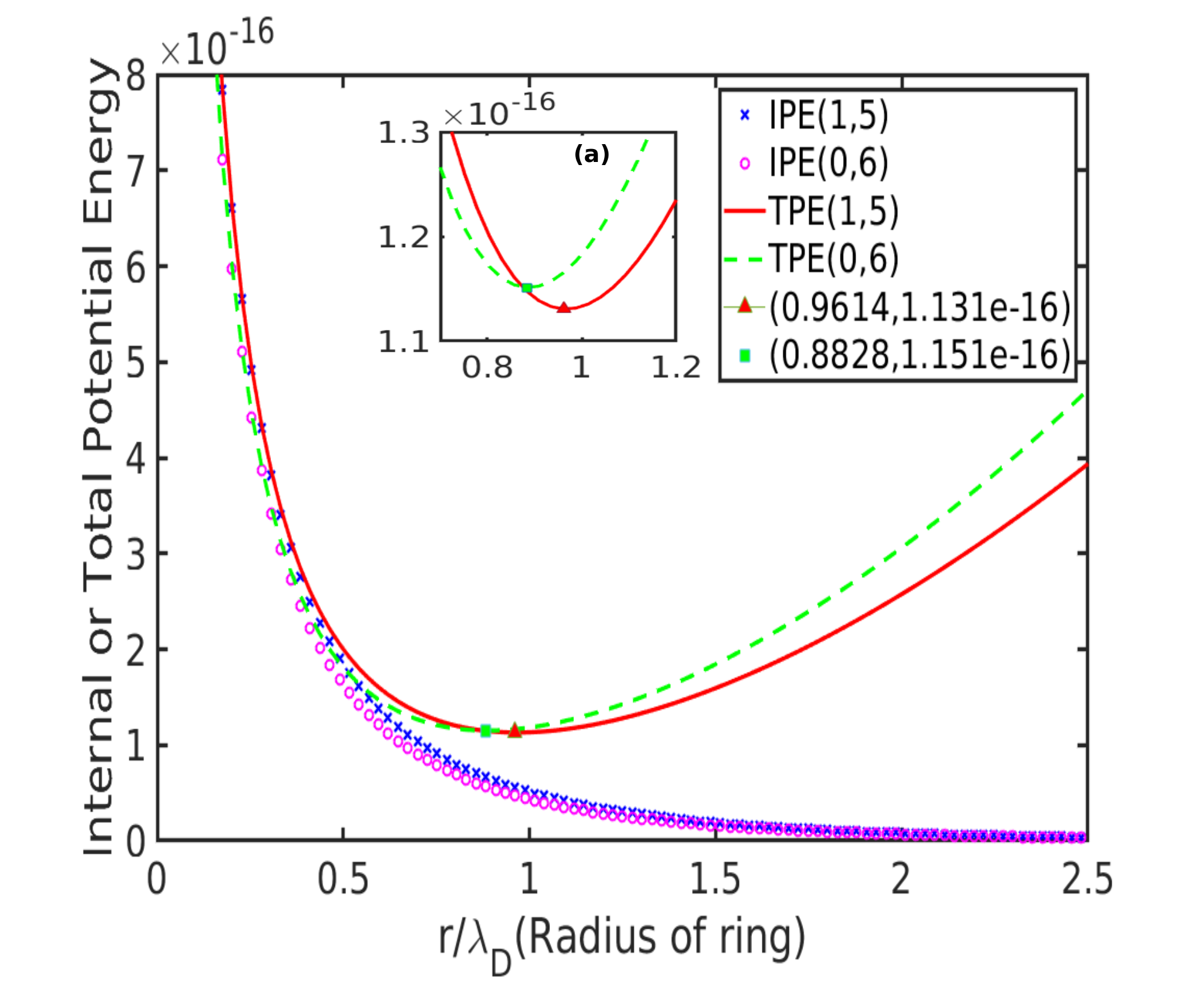}
 \caption{Plot of internal (IPE) and total potential energy (TPE) as a function of ring radius (r). Here square symbol in green and triangle in red colour represents minima of TPE of (0,6) and (1,5) respectively. Subplot (a) is the zoomed plot near minima.}
 \label{trj1}
\end{figure}

  \begin{table}[ht]
  \centering
  \includegraphics[height = 18.0cm,width = 7.0cm]{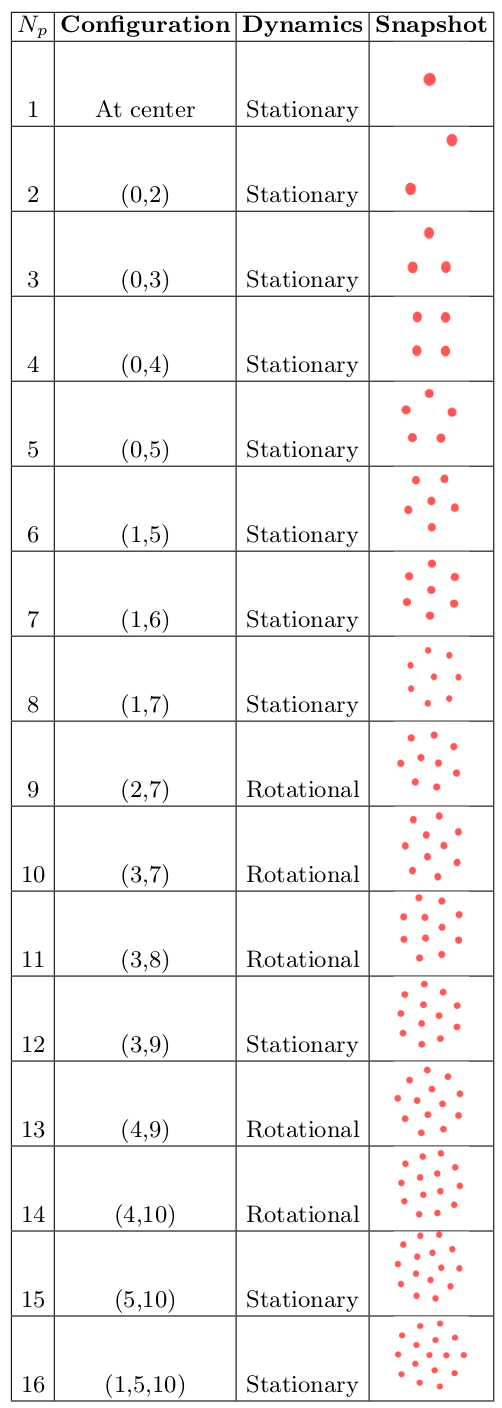}
  \caption{Table for configuration, dynamics and snapshot of shell structure at a time for different cluster systems on varying number of particles. }
 \label{table}
  \end{table}
We now try to understand the formation of these patterns.  The first question that we address is the preferred formation of the structure $(1,5)$ over $(0,6)$ 
observed in our simulations. We essentially would like to see if the former configuration is a state with lower potential energy 
for our simulations and hence it is preferred. 
 
 For this purpose we evaluate the internal potential energy (IPE) and the total potential energy (TPE)  of the two possible configurations. 
Here the  IPE is the amount of work done to bring the charges from infinite and make the cluster in the absence of an external electric field. TPE is the sum of IPE and parabolic potential energy due to external electric field. The schematic for (1,5) and (0,6) configurations is shown in Fig.\ref{confi}. 
The expression for IPE and TPE for the two   configurations  is written below: 
\begin{equation}
 IPE_{(1,5)} =\frac{5Q_d^2}{4\pi\epsilon_0}\Bigg(\frac{e^{-kr_1}}{r_1} + \frac{e^{-ka}}{a} + \frac{e^{-kb}}{b}\Bigg)
\end{equation} 
\begin{equation}
 IPE_{(0,6)} =\frac{Q_d^2}{4\pi\epsilon_0}\Bigg(\frac{6e^{-kr_2}}{r_2} + \frac{3e^{-ke}}{e} + \frac{6e^{-kd}}{d}\Bigg)
\end{equation} 
where a, b, c, d, and e can be written in terms of $r_1$ and $r_2$ using relations

\begin{align*} 
 a & = 2r_1 cos(54^{\circ})  & b & = 4r_1 cos(36^{\circ}) sin(54^{\circ}) \\  
 c & = r_2   & d & = 2r_2 cos(30^{\circ})   \\
 e & = 2r_2          \\
\end{align*}  
\begin{equation}
 TPE_{(1,5)} =IPE + \frac{5}{2}KQ_d{r_1}^2
\end{equation} 
\begin{equation}
 TPE_{(0,6)} =IPE + \frac{6}{2}KQ_d{r_1}^2
\end{equation}  
We have plotted the internal and total potential energy of both the configurations as a function of ring radius, as shown in fig.\ref{trj1}.	We find that in the absence of an external electric field, the IPE of both configurations does not have any minimum as expected. The  configurations are not stable as charged particles repel each other.  The  external electric field tries to confine the particles giving rise to a   minimum of total potential energy at a particular radius. This is  shown
 in subplot (a) of Fig.\ref{trj1}.  Here the triangle(red) and square(green) denote the minimum of TPE for two possible  configurations, viz.,  (1,5) and (0,6)  respectively. In the presence of the external electric field, the value of TPE for the  (1,5) configuration is less compared to that of (0,6) at the minima.  
 This implies that the configuration  (1,5) would be preferred over (0,6). The radius at which the minima of the  TPE occurs 
 is denoted by the red color triangle for which  $r/\lambda_D=0.9614$ shown in Fig.\ref{trj1}.  This value matches precisely with the radius of the cluster observed in the relaxed state for simulations with $6$ particles. Thus formation of these configurations is based on the  relaxation towards the minimum potential energy state. 
 
 The next question is related towards understanding those configurations which are unable to relax towards a stationary state. The characteristic feature exhibited by the dynamics in these cases is of further interest.  These issues will now be adderssed in  the next section. 
 
\section{\it Chaos in  Dynamics}
 When the total number of the particle is such that they all get accommodated in a single radial shell or with one particle at a center and others on a single 
 radial ring around it, the observed relaxed configuration is always stationary. For our simulations, this occurs till the total number of particles is $8$ in the system. 
 When we increase the number of particles, the configurations become complex.  They can be typically looked upon as structures for which particles are located 
 on two or higher numbers of shells. For these cases, if the number of particles placed on each ring is integer multiple of each other, then the system relaxes towards a stationary state. If this condition is not satisfied, then the configuration displays complex dynamics. For instance, in clusters having 10 and 11 particles with configuration $(3,7)$ and $(3,8)$ respectively are shown in Table \ref{table}. For both these cases, the particles in the two shells are not related to each other by integer numbers. It is evident that for such combinations, there exists no possible placement of particles that can ever lead to a form for which the interparticle forces acting on every particle would be get balanced by the external force field, which would be required for a stationary configuration.  
 A $\hat{\theta}$ component of force, therefore,  always remains unbalanced on some particles giving rise to interesting dynamics. 
 We now try to understand the dynamics that are exhibited by such configurations. We have specifically chosen to illustrate this 
 here with a detailed study of a  configuration with  $9$ particles. Other configurations exhibiting dynamical states have also been investigated, though they have not been presented here.  The general inferences about the dynamics for all these cases remain similar. 
 
\begin{figure}[hbt!]
   \includegraphics[height = 8.0cm,width = 9.0cm]{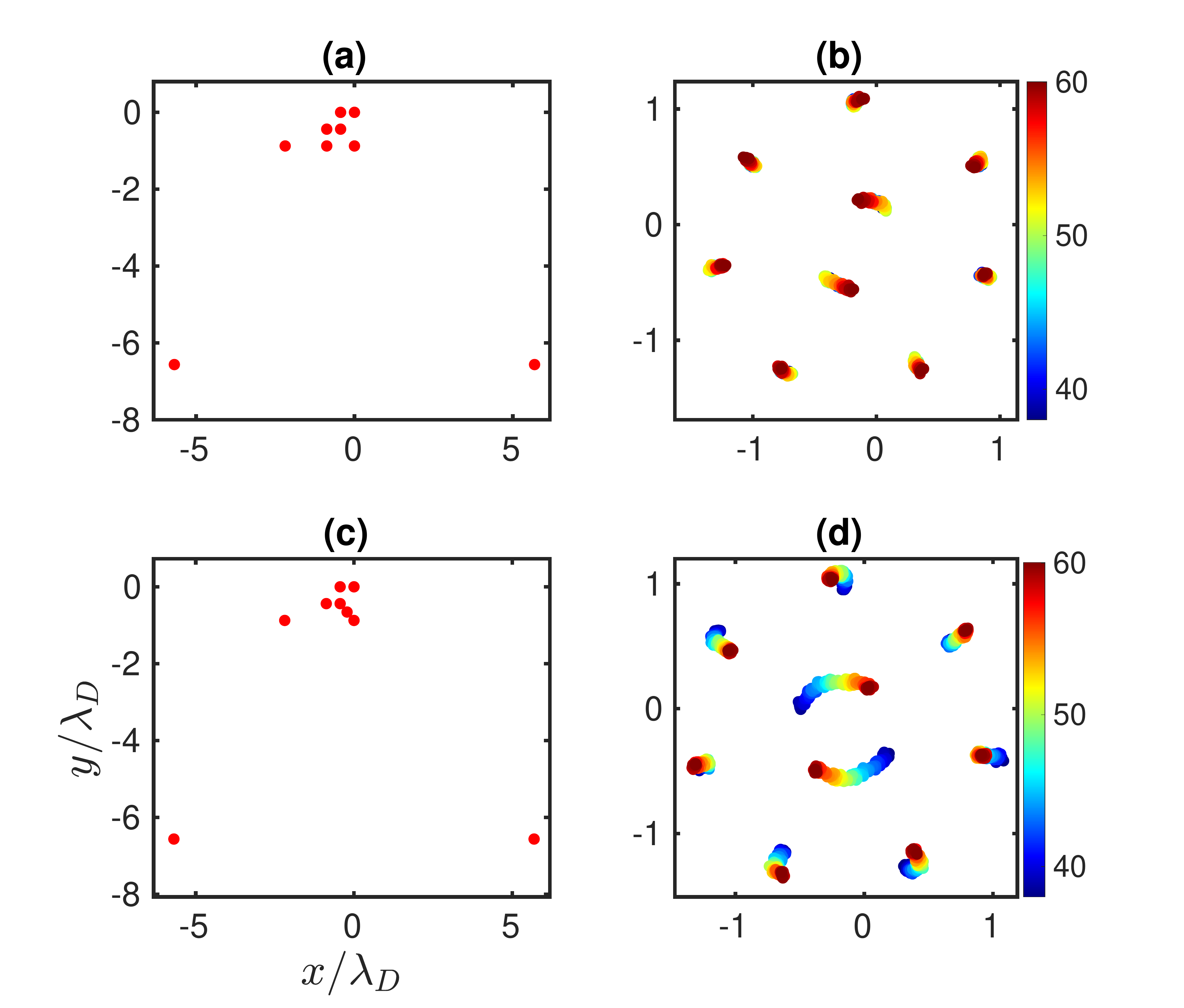}
   \caption{ Position of particles in simulation box for two different initial configurations. Subplot (a), (c) show the initial positions at time $\omega_0t $=0 and (b),(d) represent time evolution of final equilibrium configuration. Here, colorbar represnts the increase of time from the $\omega_0t $=38 to $\omega_0t $=60 }
  \label{trj3}
\end{figure}

We have shown the time evolution of the configuration of $9$ particles with two slightly different initial conditions in Fig.\ref{trj3}. In subplots (a) and (c) we have shown the particle configuration at $\omega_0t =0$. It is to be noticed that the initial configurations of these two cases are slightly differ from each other. In subplot  (b) and  (d)  the  time evolution of these two configurations have been shown from  $\omega_0t =38$ to $\omega_0t =60$. Here, the color symbols from blue to red represent the increase of time. 
  It is clear that the time evolution is drastically different even though the initial conditions were very close. The rotational dynamics observed is therefore sensitive to the chosen initial conditions. 
   
\begin{figure}[hbt!]
   \includegraphics[height = 4.0cm,width = 9.0cm]{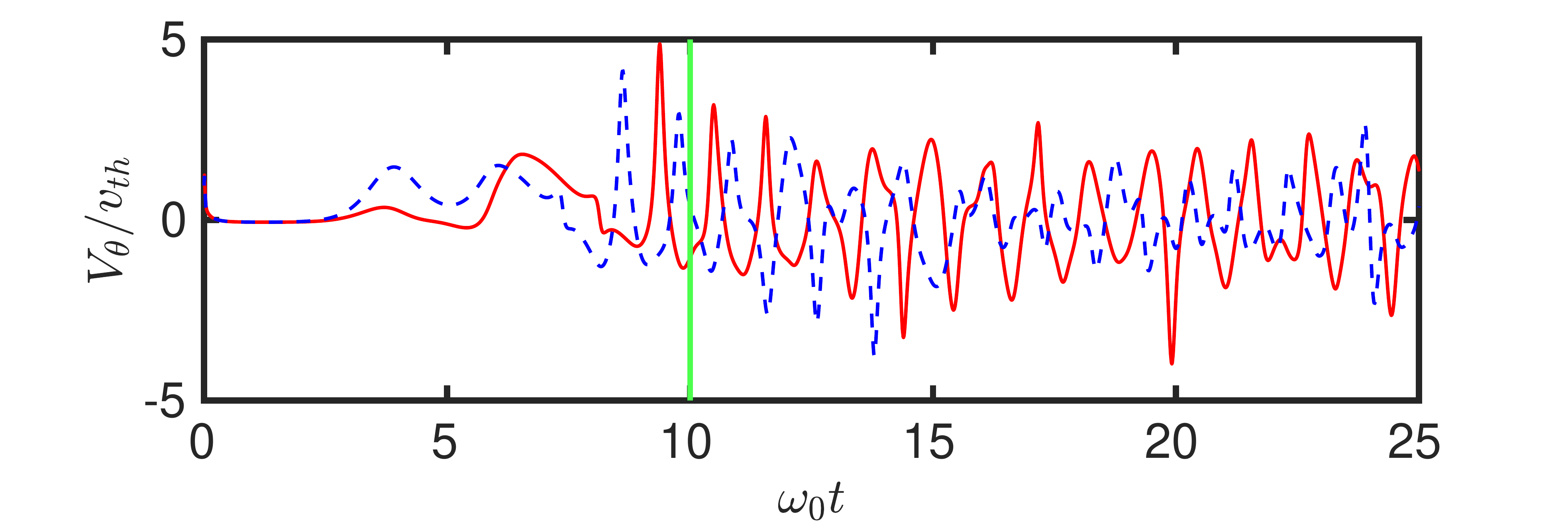}
   \caption{Time evolution of angular velocity of single particle in outer shell with slight change in initial coordinates of one particle.}
  \label{trj4}
\end{figure}

We now track the angular velocity (${V_{\theta}}$) of the particle defined by 
\begin{equation}   
      V_{\theta} = -{V_x}sin{\theta}+{V_y}cos{\theta}    
\end{equation}
    here ${V_x}$ and ${V_y}$ are the $x$ and $y$ components of particle velocity and ${\theta}$ is the angle of rotation.  In  Fig.\ref{trj4}  the time evolution ${V_{\theta}}$ has been shown for 
 one of the particle located on the  outer shells.   Here, red (solid) and blue (dash) lines represents time evolution of ${V_{\theta}}$ for two different initial space distributions of particles shown in sub-plot (a) and (c) of   Fig.\ref{trj3}.  The two plots of ${V_{\theta}}$ are almost identical to begin with and subsequently get  uncorrelated. The  vertical line in green at time $\omega_0t =10$ separates the  evolution which occurs before and after the cluster formation.  It should be observed that the changing sign of ${V_{\theta}}$ corresponds to the changing  direction of rotation.  It should be observed from the figure that the time interval at which this occurs is fairly random. 
    Thus the  system appears to be  sensitive to a slight difference in the initial condition of the particles.  We, therefore, 
    analyze this system carefully quantitatively now. 
\subsection{Time series analysis }
\begin{figure}[hbt!]
   \includegraphics[height = 8.0cm,width = 9.0cm]{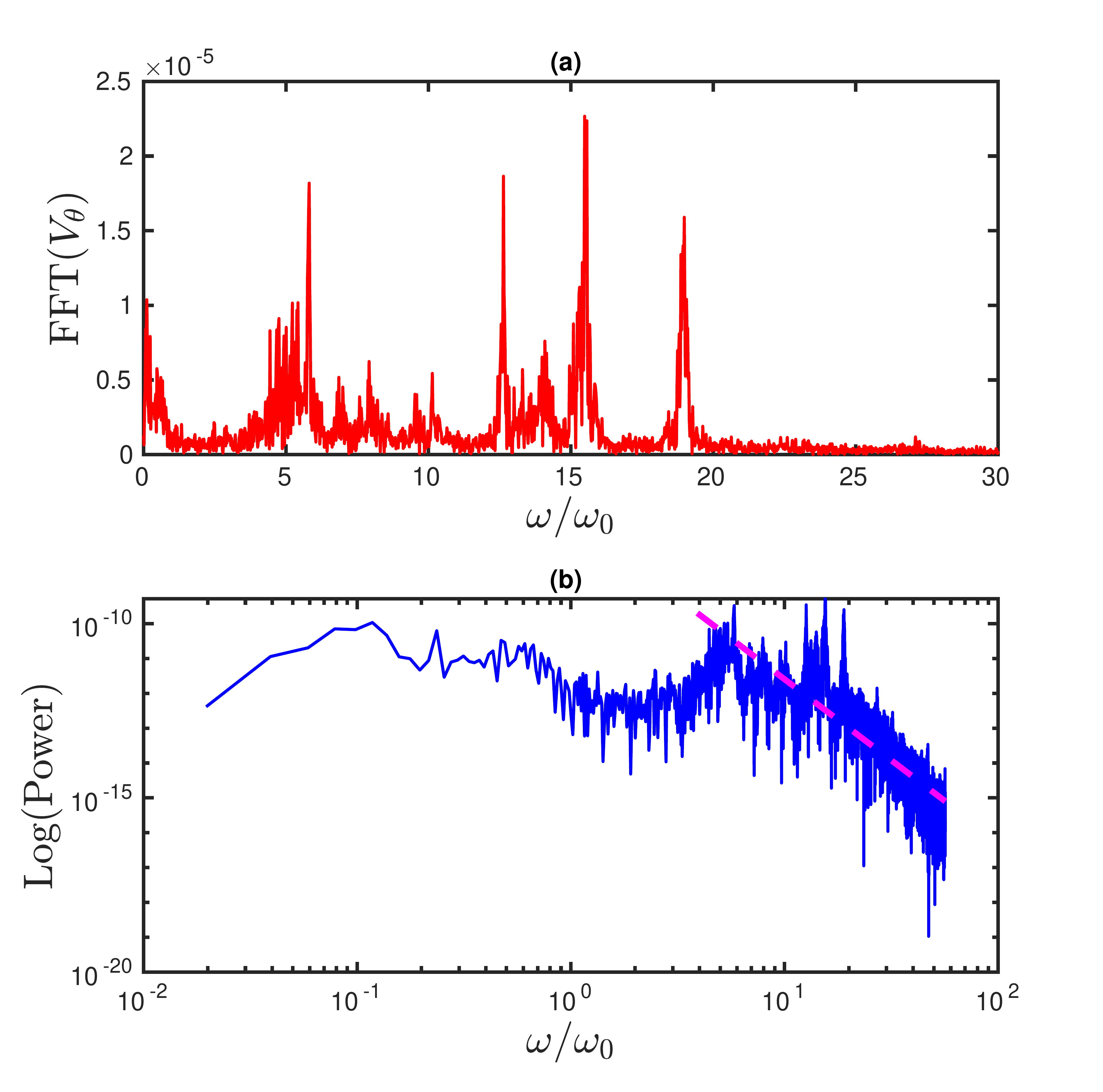}
   \caption{(a) Fourier transform (b) Power spectrum for time series of ${V_{\theta}}$, Here magenta (dash) line represents the linearly fitted slope. }
    \label{trj5}
\end{figure}

\begin{figure}[hbt!]
   \includegraphics[height = 4.0cm,width = 9.0cm]{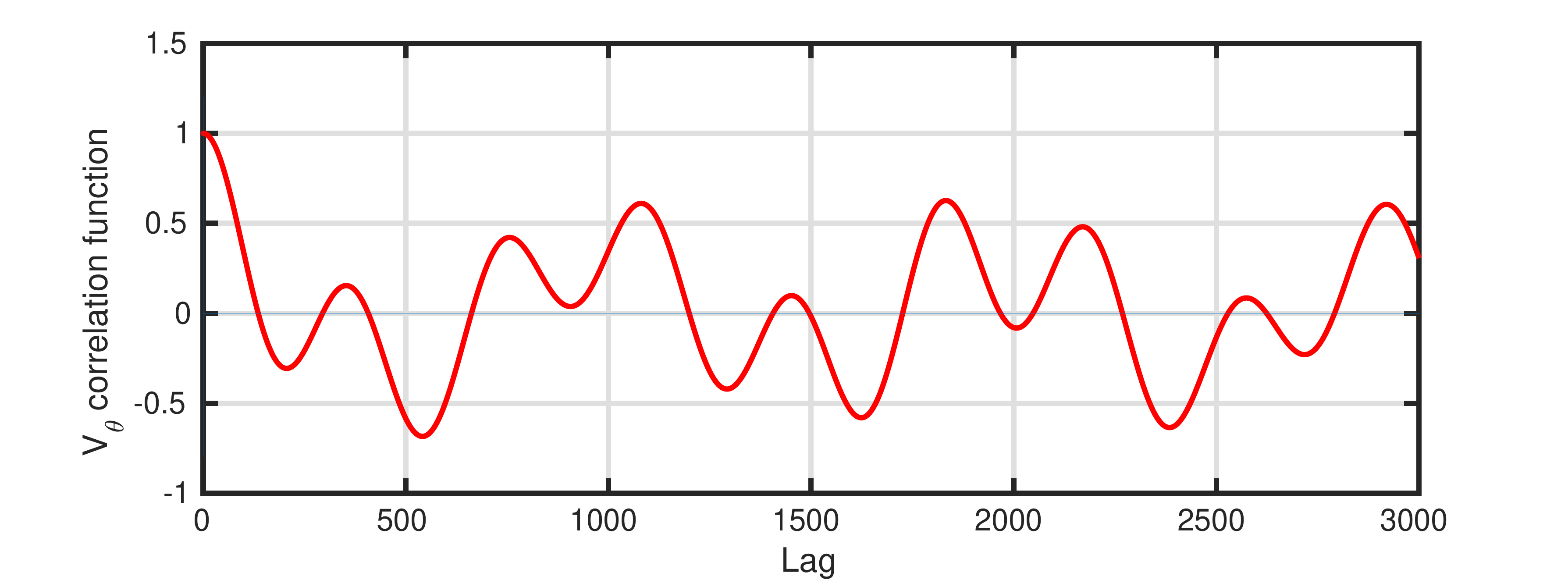}
   \caption{ Velocity auto correlation of one of the particle in inner shell of cluster.}
   \label{trj6}
\end{figure}

\begin{figure}[hbt!]
   \includegraphics[height = 7.0cm,width = 9.0cm]{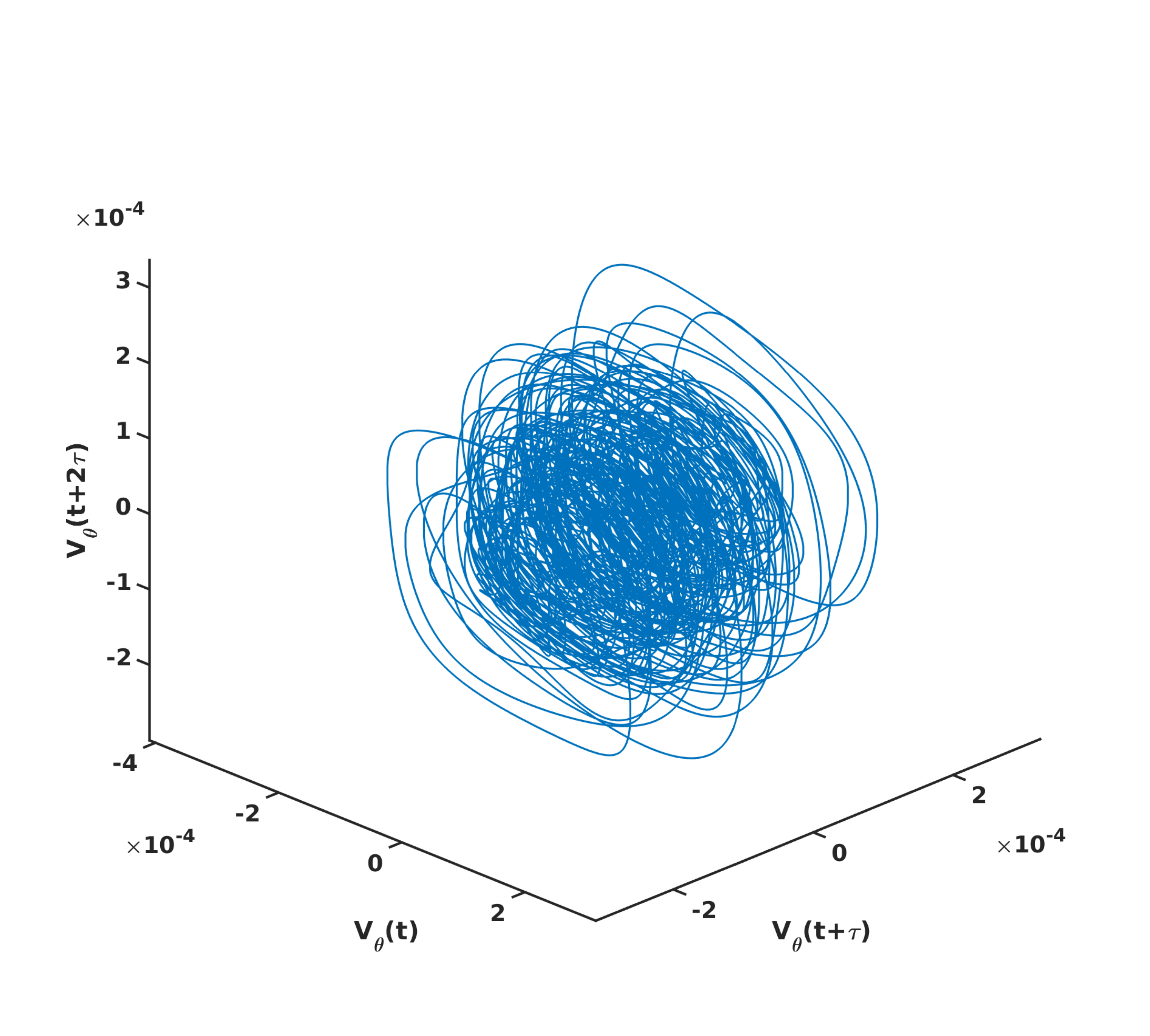}
   \caption{ Reconstructed phase space attractor in 3-D using ${V_{\theta}}$(t) with time delay of 132 time steps.}
    \label{trj7}
\end{figure}

 In this section, we will mainly focus on the time history of ${V_{\theta}}$ for one of the particles, which has shown evidence of sensitivity towards initial conditions.  In Fig.\ref{trj5} sub-plot (a) and (b) the  Fourier transform and power spectrum of ${V_{\theta}}$  from   $\omega_0t = 10$ to $666$ has been shown respectively.  
 There are definite peaks in the frequency spectra against a noisy background. We can see from these subplots that the power spectrum is considerably broad, and there is no specific characteristic frequency of the system.  The nature of the frequency and power spectrum is broad.  The sub-plot (b) of Fig.\ref{trj5} shows that at the higher end of the spectrum, the power spectrum of ${V_{\theta}}$ falls as  $\omega^{-4}$.
  Identification of chaos in a series is a multistage process that includes calculating time delay coordinate, reconstructing phase space, calculating correlation dimension, and Lyapunov exponent \cite{baker1996chaotic}.  The first step constitutes calculating the time-delay coordinate ($\tau$).  We want to reconstruct a phase space attractor, so it's important to get a good estimate for the time delay ($\tau$) for this time series.  The value of $\tau$ is a typical time after which one expects the correlation in the signal to die out. There are several methods by which we can calculate this time lag \cite{fraser1986independent,martinerie1992mutual,fraser1989information}.
  It can thus be calculated by using either velocity autocorrelation or mutual information function. Fraser and Swinney \cite{fraser1986independent} introduced an approach for selecting time delay using the first local minimum of mutual information function. Velocity auto-correlation examines the correlation in time series data as a function of time, and the first zero crossing gave an idea about $\tau$. Here, we have a time-series of ${V_{\theta}}$ and its auto-correlation is shown in Fig.\ref{trj6}. The first zero crossing corresponds to a delay of $136$-time steps, which gives us time delay $\tau= 0.150\omega_0t$. Choosing this method would not introduce any bias as invariant quantities computing using reconstructed attractor are not very much sensitive to $\tau$ \cite{baker1996chaotic}. 
	The second step constitutes reconstruction of the phase space for this attractor, an abstract mathematical space spanned by the dynamical variable of the system.  It was shown by Takens \cite{takens1981detecting} that phase space can be reconstructed by time-delayed measurement of a single observed time-series signal.   Fig.\ref{trj7} shows a reconstructed 3-D attractor for the time series of ${V_{\theta}}$. In order to resolve the structure of the system in reconstructed phase space, the minimum embedding dimension $m$ is found to be $3$ using ``False Nearest Neighbour'' \cite{kennel1992determining}. For the choice of $m=3$ there were no self-intersections. 
    The third step is to calculate correlation dimension \cite{grassberger1983characterization} $(d)$.  It should be independent of embedding dimension $m$, which gives information about attractor that is effectively embedded in higher dimension space.  We found that correlation dimension $d$ is $2.25 \pm 0.05$.  We observe that on changing the embedding dimension, the correlation dimension becomes independent of $m$ as shown in Fig.8.  This suggests that the particle dynamics is not totally random but has chaotic attribute for which the attractor is strange with a non-integral dimension. 
  \begin{figure}[hbt!]
   \includegraphics[height = 5.0cm,width = 9.0cm]{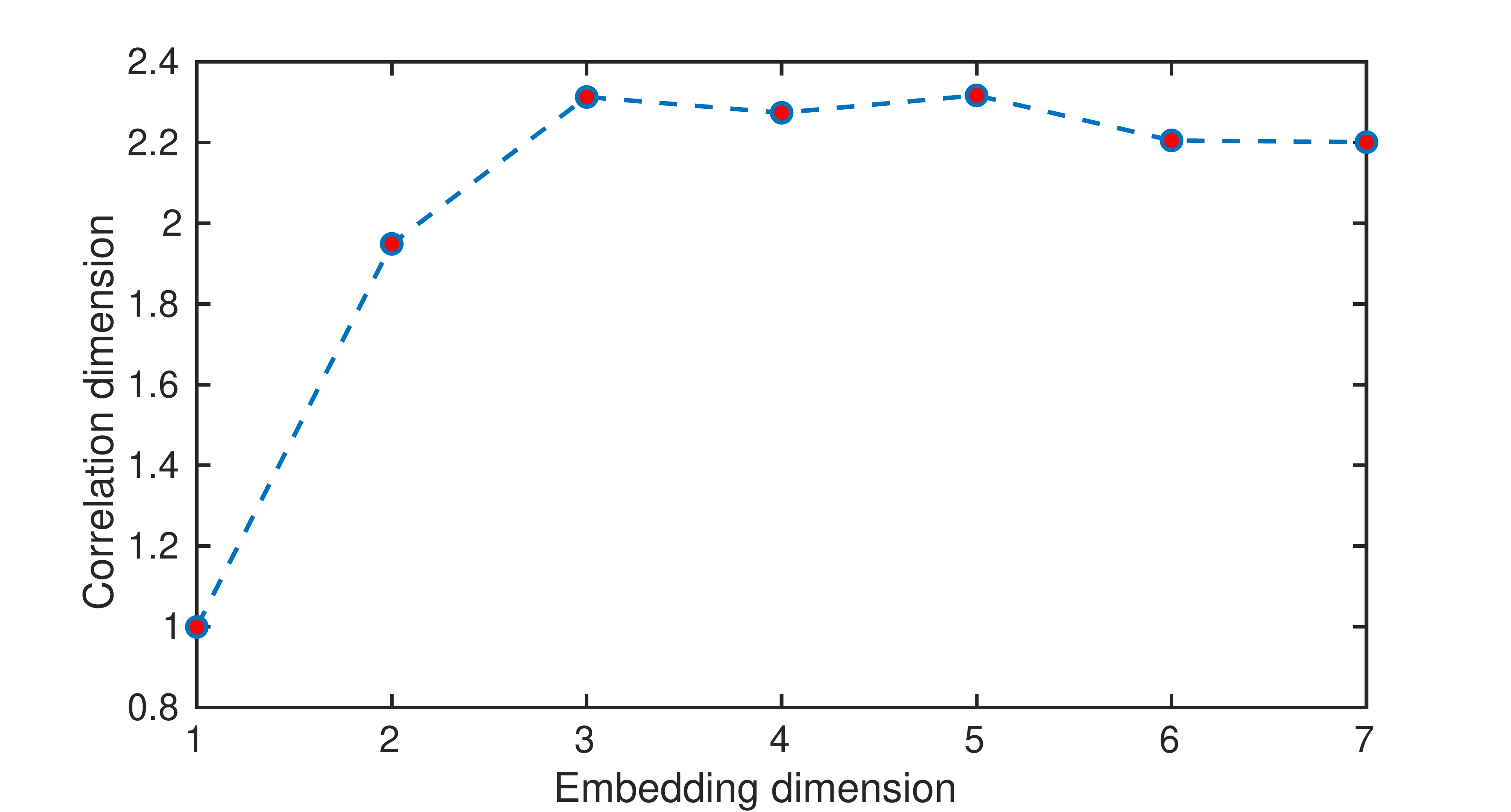}
   \caption{ Embedding dimension Vs correlation dimension. }
    \label{trj8}
\end{figure}  
	As the final step of our analysis, we try to find the  Lyapunov exponent, which quantifies the mean divergence between neighboring trajectories in 
	phase space for a chaotic system.  The chaotic system must have at least one non-negative Lyapunov exponent.  We evaluated the largest Lyapunov exponent (LLE) using the Rosenstein algorithm \cite{rosenstein1993practical} by allocating the nearest neighbor on adjacent trajectories and computing the divergence between successive pairs along the trajectories. The slope of average logarithmic divergence is shown in subplot(a) of Fig.\ref{trj9} which gives the value of LLE as $ 0.21496$. The positive value of the LLE confirms that the system is chaotic. 
	
	We have carried out a similar analysis for other cluster configurations and observed that the rotational dynamics exhibited by the cluster configuration is essentially chaotic in nature. The Lyapunov index and other characteristics of the attractor for each case may, however, differ. For instance, in a cluster having $13$ particles the largest Lyapunov exponent is $0.16898$ as shown in subplot (b) of Fig.\ref{trj9}. The observation of strange attractors and positive LLE confirms the chaotic dynamics of particles inside the cluster.

	\begin{figure}[hbt!]
   \includegraphics[height = 7.0cm,width = 7.5cm]{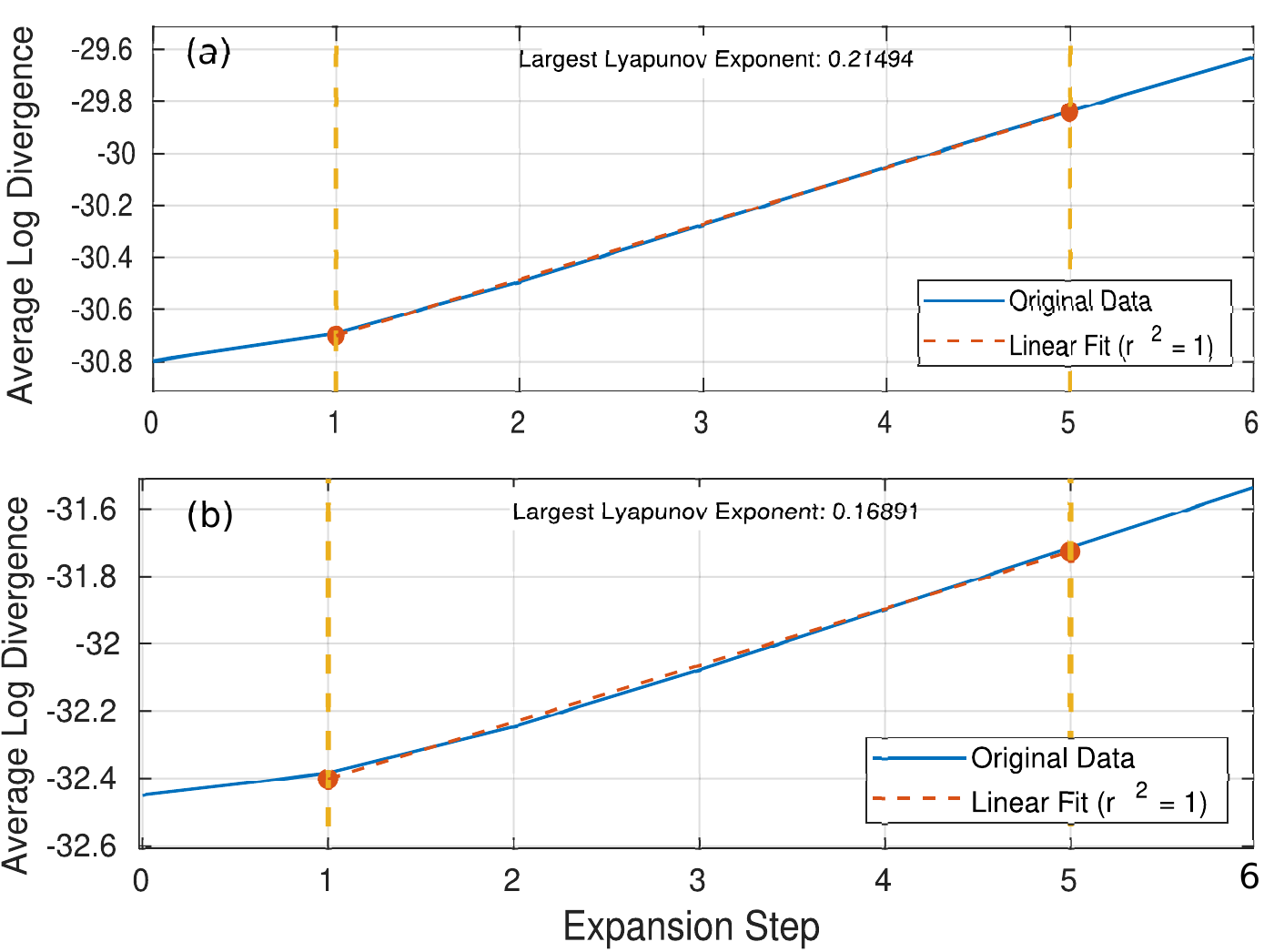}
   \caption{Lyapunov exponent for cluster having (a) 9 and (b) 13 particles.  }
    \label{trj9}
\end{figure} 
	We have also carried out Langevin dynamics simulation \cite{feng2010identifying} using LAMMPS \cite{plimpton1995fast}, which includes the effect of neutral on dust particles of the cluster. The phase space trajectories in the attractor are a little distorted due to random kicks of neutrals with dust grains, but dynamics still remain chaotic with small changes in parameters like correlation dimension and Lyapunov exponent.


\section {Conclusion} 
Equilibrium and/or relaxed state for particles interacting with repulsive screened Coulomb/Yukawa potential in an overall radially confining force field was studied using Molecular Dynamics simulation. Such a system can be prepared easily in the laboratory by immersing charged micron-sized dust particles in ordinary electron-ion plasma. A biased ring electrode can provide the radial confinement.  It is observed that the system relaxes towards a minimum energy configuration. In such a state, particles organize themselves in various shells around the center. 
Depending on the number of particles, the relaxed state is observed to be either stationary or exhibits a dynamical rotating state in which the particles 
arranged in various shells show rotation. The rotation changes with time, and detailed analysis shows that the dynamics is chaotic. 

\section{Acknowledgements}
This research work has been supported by the
 J. C. Bose fellowship grant of AD (JCB/2017/000055) and
the CRG/2018/000624 grant of DST. The authors thank IIT Delhi HPC facility for computational resources.
\newpage

\bibliography{cluster_ref}

\begin{thebibliography}{31}
\expandafter\ifx\csname natexlab\endcsname\relax\def\natexlab#1{#1}\fi
\expandafter\ifx\csname bibnamefont\endcsname\relax
  \def\bibnamefont#1{#1}\fi
\expandafter\ifx\csname bibfnamefont\endcsname\relax
  \def\bibfnamefont#1{#1}\fi
\expandafter\ifx\csname citenamefont\endcsname\relax
  \def\citenamefont#1{#1}\fi
\expandafter\ifx\csname url\endcsname\relax
  \def\url#1{\texttt{#1}}\fi
\expandafter\ifx\csname urlprefix\endcsname\relax\def\urlprefix{URL }\fi
\providecommand{\bibinfo}[2]{#2}
\providecommand{\eprint}[2][]{\url{#2}}

\bibitem[{\citenamefont{Maity et~al.}(2020)\citenamefont{Maity, Deshwal, Yadav,
  and Das}}]{maity2020dynamical}
\bibinfo{author}{\bibfnamefont{S.}~\bibnamefont{Maity}},
  \bibinfo{author}{\bibfnamefont{P.}~\bibnamefont{Deshwal}},
  \bibinfo{author}{\bibfnamefont{M.}~\bibnamefont{Yadav}}, \bibnamefont{and}
  \bibinfo{author}{\bibfnamefont{A.}~\bibnamefont{Das}},
  \bibinfo{journal}{Physical Review E} \textbf{\bibinfo{volume}{102}},
  \bibinfo{pages}{023213} (\bibinfo{year}{2020}).

\bibitem[{\citenamefont{Fan et~al.}(1992)\citenamefont{Fan, Yang, Dai, Zheng,
  Yuan, and Tsai}}]{fan1992observations}
\bibinfo{author}{\bibfnamefont{S.}~\bibnamefont{Fan}},
  \bibinfo{author}{\bibfnamefont{S.}~\bibnamefont{Yang}},
  \bibinfo{author}{\bibfnamefont{J.}~\bibnamefont{Dai}},
  \bibinfo{author}{\bibfnamefont{S.}~\bibnamefont{Zheng}},
  \bibinfo{author}{\bibfnamefont{D.}~\bibnamefont{Yuan}}, \bibnamefont{and}
  \bibinfo{author}{\bibfnamefont{S.}~\bibnamefont{Tsai}},
  \bibinfo{journal}{Physics Letters A} \textbf{\bibinfo{volume}{164}},
  \bibinfo{pages}{295} (\bibinfo{year}{1992}).

\bibitem[{\citenamefont{Mitra et~al.}(2014)\citenamefont{Mitra, Sarma, Janaki,
  Iyenger, Sarma, Marwan, Kurths, Shaw, Saha, and Ghosh}}]{mitra2014order}
\bibinfo{author}{\bibfnamefont{V.}~\bibnamefont{Mitra}},
  \bibinfo{author}{\bibfnamefont{A.}~\bibnamefont{Sarma}},
  \bibinfo{author}{\bibfnamefont{M.}~\bibnamefont{Janaki}},
  \bibinfo{author}{\bibfnamefont{A.~S.} \bibnamefont{Iyenger}},
  \bibinfo{author}{\bibfnamefont{B.}~\bibnamefont{Sarma}},
  \bibinfo{author}{\bibfnamefont{N.}~\bibnamefont{Marwan}},
  \bibinfo{author}{\bibfnamefont{J.}~\bibnamefont{Kurths}},
  \bibinfo{author}{\bibfnamefont{P.~K.} \bibnamefont{Shaw}},
  \bibinfo{author}{\bibfnamefont{D.}~\bibnamefont{Saha}}, \bibnamefont{and}
  \bibinfo{author}{\bibfnamefont{S.}~\bibnamefont{Ghosh}},
  \bibinfo{journal}{Chaos, Solitons \& Fractals} \textbf{\bibinfo{volume}{69}},
  \bibinfo{pages}{285} (\bibinfo{year}{2014}).

\bibitem[{\citenamefont{Shaw et~al.}(2019)\citenamefont{Shaw, Chaubey,
  Mukherjee, Janaki, and Iyengar}}]{shaw2019continuous}
\bibinfo{author}{\bibfnamefont{P.~K.} \bibnamefont{Shaw}},
  \bibinfo{author}{\bibfnamefont{N.}~\bibnamefont{Chaubey}},
  \bibinfo{author}{\bibfnamefont{S.}~\bibnamefont{Mukherjee}},
  \bibinfo{author}{\bibfnamefont{M.}~\bibnamefont{Janaki}}, \bibnamefont{and}
  \bibinfo{author}{\bibfnamefont{A.~S.} \bibnamefont{Iyengar}},
  \bibinfo{journal}{Physica A: Statistical Mechanics and its Applications}
  \textbf{\bibinfo{volume}{513}}, \bibinfo{pages}{126} (\bibinfo{year}{2019}).

\bibitem[{\citenamefont{Sheridan}(2005)}]{sheridan2005chaos}
\bibinfo{author}{\bibfnamefont{T.}~\bibnamefont{Sheridan}},
  \bibinfo{journal}{Physics of plasmas} \textbf{\bibinfo{volume}{12}},
  \bibinfo{pages}{080701} (\bibinfo{year}{2005}).

\bibitem[{\citenamefont{Sheridan and Theisen}(2010)}]{sheridan2010transition}
\bibinfo{author}{\bibfnamefont{T.}~\bibnamefont{Sheridan}} \bibnamefont{and}
  \bibinfo{author}{\bibfnamefont{W.}~\bibnamefont{Theisen}},
  \bibinfo{journal}{Physics of Plasmas} \textbf{\bibinfo{volume}{17}},
  \bibinfo{pages}{013703} (\bibinfo{year}{2010}).

\bibitem[{\citenamefont{Juan et~al.}(1998)\citenamefont{Juan, Huang, Hsu, Lai,
  and Lin}}]{juan1998observation}
\bibinfo{author}{\bibfnamefont{W.-T.} \bibnamefont{Juan}},
  \bibinfo{author}{\bibfnamefont{Z.-H.} \bibnamefont{Huang}},
  \bibinfo{author}{\bibfnamefont{J.-W.} \bibnamefont{Hsu}},
  \bibinfo{author}{\bibfnamefont{Y.-J.} \bibnamefont{Lai}}, \bibnamefont{and}
  \bibinfo{author}{\bibfnamefont{I.}~\bibnamefont{Lin}},
  \bibinfo{journal}{Physical Review E} \textbf{\bibinfo{volume}{58}},
  \bibinfo{pages}{R6947} (\bibinfo{year}{1998}).

\bibitem[{\citenamefont{Murillo}(2004)}]{murillo2004strongly}
\bibinfo{author}{\bibfnamefont{M.~S.} \bibnamefont{Murillo}},
  \bibinfo{journal}{Physics of Plasmas} \textbf{\bibinfo{volume}{11}},
  \bibinfo{pages}{2964} (\bibinfo{year}{2004}).

\bibitem[{\citenamefont{Chu and Lin}(1994)}]{chu1994direct}
\bibinfo{author}{\bibfnamefont{J.}~\bibnamefont{Chu}} \bibnamefont{and}
  \bibinfo{author}{\bibfnamefont{I.}~\bibnamefont{Lin}},
  \bibinfo{journal}{Physical review letters} \textbf{\bibinfo{volume}{72}},
  \bibinfo{pages}{4009} (\bibinfo{year}{1994}).

\bibitem[{\citenamefont{Thomas et~al.}(1994)\citenamefont{Thomas, Morfill,
  Demmel, Goree, Feuerbacher, and M{\"o}hlmann}}]{thomas1994plasma}
\bibinfo{author}{\bibfnamefont{H.}~\bibnamefont{Thomas}},
  \bibinfo{author}{\bibfnamefont{G.}~\bibnamefont{Morfill}},
  \bibinfo{author}{\bibfnamefont{V.}~\bibnamefont{Demmel}},
  \bibinfo{author}{\bibfnamefont{J.}~\bibnamefont{Goree}},
  \bibinfo{author}{\bibfnamefont{B.}~\bibnamefont{Feuerbacher}},
  \bibnamefont{and}
  \bibinfo{author}{\bibfnamefont{D.}~\bibnamefont{M{\"o}hlmann}},
  \bibinfo{journal}{Physical Review Letters} \textbf{\bibinfo{volume}{73}},
  \bibinfo{pages}{652} (\bibinfo{year}{1994}).

\bibitem[{\citenamefont{Hayashi and Tachibana}(1994)}]{hayashi1994observation}
\bibinfo{author}{\bibfnamefont{Y.}~\bibnamefont{Hayashi}} \bibnamefont{and}
  \bibinfo{author}{\bibfnamefont{K.}~\bibnamefont{Tachibana}},
  \bibinfo{journal}{Jpn. J. Appl. Phys.} \textbf{\bibinfo{volume}{33}},
  \bibinfo{pages}{L804} (\bibinfo{year}{1994}).

\bibitem[{\citenamefont{Shukla and Mamun}(2015)}]{shukla2015introduction}
\bibinfo{author}{\bibfnamefont{P.~K.} \bibnamefont{Shukla}} \bibnamefont{and}
  \bibinfo{author}{\bibfnamefont{A.}~\bibnamefont{Mamun}},
  \emph{\bibinfo{title}{Introduction to dusty plasma physics}}
  (\bibinfo{publisher}{CRC press}, \bibinfo{year}{2015}).

\bibitem[{\citenamefont{Juan and Lin}(1998)}]{juan1998anomalous}
\bibinfo{author}{\bibfnamefont{W.-T.} \bibnamefont{Juan}} \bibnamefont{and}
  \bibinfo{author}{\bibfnamefont{I.}~\bibnamefont{Lin}},
  \bibinfo{journal}{Physical review letters} \textbf{\bibinfo{volume}{80}},
  \bibinfo{pages}{3073} (\bibinfo{year}{1998}).

\bibitem[{\citenamefont{Teng et~al.}(2009)\citenamefont{Teng, Chang, Tseng, and
  Lin}}]{teng2009wave}
\bibinfo{author}{\bibfnamefont{L.-W.} \bibnamefont{Teng}},
  \bibinfo{author}{\bibfnamefont{M.-C.} \bibnamefont{Chang}},
  \bibinfo{author}{\bibfnamefont{Y.-P.} \bibnamefont{Tseng}}, \bibnamefont{and}
  \bibinfo{author}{\bibfnamefont{I.}~\bibnamefont{Lin}},
  \bibinfo{journal}{Physical review letters} \textbf{\bibinfo{volume}{103}},
  \bibinfo{pages}{245005} (\bibinfo{year}{2009}).

\bibitem[{\citenamefont{Maity et~al.}(2018)\citenamefont{Maity, Das, Kumar, and
  Tiwari}}]{maity2018interplay}
\bibinfo{author}{\bibfnamefont{S.}~\bibnamefont{Maity}},
  \bibinfo{author}{\bibfnamefont{A.}~\bibnamefont{Das}},
  \bibinfo{author}{\bibfnamefont{S.}~\bibnamefont{Kumar}}, \bibnamefont{and}
  \bibinfo{author}{\bibfnamefont{S.~K.} \bibnamefont{Tiwari}},
  \bibinfo{journal}{Physics of Plasmas} \textbf{\bibinfo{volume}{25}},
  \bibinfo{pages}{043705} (\bibinfo{year}{2018}).

\bibitem[{\citenamefont{Melzer et~al.}(1996)\citenamefont{Melzer, Homann, and
  Piel}}]{melzer1996experimental}
\bibinfo{author}{\bibfnamefont{A.}~\bibnamefont{Melzer}},
  \bibinfo{author}{\bibfnamefont{A.}~\bibnamefont{Homann}}, \bibnamefont{and}
  \bibinfo{author}{\bibfnamefont{A.}~\bibnamefont{Piel}},
  \bibinfo{journal}{Physical Review E} \textbf{\bibinfo{volume}{53}},
  \bibinfo{pages}{2757} (\bibinfo{year}{1996}).

\bibitem[{\citenamefont{Schweigert et~al.}(1998)\citenamefont{Schweigert,
  Schweigert, Melzer, Homann, and Piel}}]{schweigert1998plasma}
\bibinfo{author}{\bibfnamefont{V.}~\bibnamefont{Schweigert}},
  \bibinfo{author}{\bibfnamefont{I.}~\bibnamefont{Schweigert}},
  \bibinfo{author}{\bibfnamefont{A.}~\bibnamefont{Melzer}},
  \bibinfo{author}{\bibfnamefont{A.}~\bibnamefont{Homann}}, \bibnamefont{and}
  \bibinfo{author}{\bibfnamefont{A.}~\bibnamefont{Piel}},
  \bibinfo{journal}{Physical review letters} \textbf{\bibinfo{volume}{80}},
  \bibinfo{pages}{5345} (\bibinfo{year}{1998}).

\bibitem[{\citenamefont{Maity and Das}(2019)}]{maity2019molecular}
\bibinfo{author}{\bibfnamefont{S.}~\bibnamefont{Maity}} \bibnamefont{and}
  \bibinfo{author}{\bibfnamefont{A.}~\bibnamefont{Das}},
  \bibinfo{journal}{Physics of Plasmas} \textbf{\bibinfo{volume}{26}},
  \bibinfo{pages}{023703} (\bibinfo{year}{2019}).

\bibitem[{\citenamefont{Plimpton}(1995)}]{plimpton1995fast}
\bibinfo{author}{\bibfnamefont{S.}~\bibnamefont{Plimpton}},
  \bibinfo{journal}{Journal of computational physics}
  \textbf{\bibinfo{volume}{117}}, \bibinfo{pages}{1} (\bibinfo{year}{1995}).

\bibitem[{\citenamefont{Nosenko and Goree}(2004)}]{nosenko2004shear}
\bibinfo{author}{\bibfnamefont{V.}~\bibnamefont{Nosenko}} \bibnamefont{and}
  \bibinfo{author}{\bibfnamefont{J.}~\bibnamefont{Goree}},
  \bibinfo{journal}{Physical review letters} \textbf{\bibinfo{volume}{93}},
  \bibinfo{pages}{155004} (\bibinfo{year}{2004}).

\bibitem[{\citenamefont{Nos{\'e}}(1984)}]{nose1984molecular}
\bibinfo{author}{\bibfnamefont{S.}~\bibnamefont{Nos{\'e}}},
  \bibinfo{journal}{Molecular physics} \textbf{\bibinfo{volume}{52}},
  \bibinfo{pages}{255} (\bibinfo{year}{1984}).

\bibitem[{\citenamefont{Hoover}(1985)}]{hoover1985canonical}
\bibinfo{author}{\bibfnamefont{W.~G.} \bibnamefont{Hoover}},
  \bibinfo{journal}{Physical review A} \textbf{\bibinfo{volume}{31}},
  \bibinfo{pages}{1695} (\bibinfo{year}{1985}).

\bibitem[{\citenamefont{Baker and Gollub}(1996)}]{baker1996chaotic}
\bibinfo{author}{\bibfnamefont{G.~L.} \bibnamefont{Baker}} \bibnamefont{and}
  \bibinfo{author}{\bibfnamefont{J.~P.} \bibnamefont{Gollub}},
  \emph{\bibinfo{title}{Chaotic dynamics: an introduction}}
  (\bibinfo{publisher}{Cambridge university press}, \bibinfo{year}{1996}).

\bibitem[{\citenamefont{Fraser and Swinney}(1986)}]{fraser1986independent}
\bibinfo{author}{\bibfnamefont{A.~M.} \bibnamefont{Fraser}} \bibnamefont{and}
  \bibinfo{author}{\bibfnamefont{H.~L.} \bibnamefont{Swinney}},
  \bibinfo{journal}{Physical review A} \textbf{\bibinfo{volume}{33}},
  \bibinfo{pages}{1134} (\bibinfo{year}{1986}).

\bibitem[{\citenamefont{Martinerie et~al.}(1992)\citenamefont{Martinerie,
  Albano, Mees, and Rapp}}]{martinerie1992mutual}
\bibinfo{author}{\bibfnamefont{J.}~\bibnamefont{Martinerie}},
  \bibinfo{author}{\bibfnamefont{A.~M.} \bibnamefont{Albano}},
  \bibinfo{author}{\bibfnamefont{A.}~\bibnamefont{Mees}}, \bibnamefont{and}
  \bibinfo{author}{\bibfnamefont{P.}~\bibnamefont{Rapp}},
  \bibinfo{journal}{Physical Review A} \textbf{\bibinfo{volume}{45}},
  \bibinfo{pages}{7058} (\bibinfo{year}{1992}).

\bibitem[{\citenamefont{Fraser}(1989)}]{fraser1989information}
\bibinfo{author}{\bibfnamefont{A.~M.} \bibnamefont{Fraser}},
  \bibinfo{journal}{IEEE transactions on Information Theory}
  \textbf{\bibinfo{volume}{35}}, \bibinfo{pages}{245} (\bibinfo{year}{1989}).

\bibitem[{\citenamefont{Takens}(1981)}]{takens1981detecting}
\bibinfo{author}{\bibfnamefont{F.}~\bibnamefont{Takens}}, in
  \emph{\bibinfo{booktitle}{Dynamical systems and turbulence, Warwick 1980}}
  (\bibinfo{publisher}{Springer}, \bibinfo{year}{1981}), pp.
  \bibinfo{pages}{366--381}.

\bibitem[{\citenamefont{Kennel et~al.}(1992)\citenamefont{Kennel, Brown, and
  Abarbanel}}]{kennel1992determining}
\bibinfo{author}{\bibfnamefont{M.~B.} \bibnamefont{Kennel}},
  \bibinfo{author}{\bibfnamefont{R.}~\bibnamefont{Brown}}, \bibnamefont{and}
  \bibinfo{author}{\bibfnamefont{H.~D.} \bibnamefont{Abarbanel}},
  \bibinfo{journal}{Physical review A} \textbf{\bibinfo{volume}{45}},
  \bibinfo{pages}{3403} (\bibinfo{year}{1992}).

\bibitem[{\citenamefont{Grassberger and
  Procaccia}(1983)}]{grassberger1983characterization}
\bibinfo{author}{\bibfnamefont{P.}~\bibnamefont{Grassberger}} \bibnamefont{and}
  \bibinfo{author}{\bibfnamefont{I.}~\bibnamefont{Procaccia}},
  \bibinfo{journal}{Physical review letters} \textbf{\bibinfo{volume}{50}},
  \bibinfo{pages}{346} (\bibinfo{year}{1983}).

\bibitem[{\citenamefont{Rosenstein et~al.}(1993)\citenamefont{Rosenstein,
  Collins, and De~Luca}}]{rosenstein1993practical}
\bibinfo{author}{\bibfnamefont{M.~T.} \bibnamefont{Rosenstein}},
  \bibinfo{author}{\bibfnamefont{J.~J.} \bibnamefont{Collins}},
  \bibnamefont{and} \bibinfo{author}{\bibfnamefont{C.~J.}
  \bibnamefont{De~Luca}}, \bibinfo{journal}{Physica D: Nonlinear Phenomena}
  \textbf{\bibinfo{volume}{65}}, \bibinfo{pages}{117} (\bibinfo{year}{1993}).

\bibitem[{\citenamefont{Feng et~al.}(2010)\citenamefont{Feng, Goree, and
  Liu}}]{feng2010identifying}
\bibinfo{author}{\bibfnamefont{Y.}~\bibnamefont{Feng}},
  \bibinfo{author}{\bibfnamefont{J.}~\bibnamefont{Goree}}, \bibnamefont{and}
  \bibinfo{author}{\bibfnamefont{B.}~\bibnamefont{Liu}},
  \bibinfo{journal}{Physical Review E} \textbf{\bibinfo{volume}{82}},
  \bibinfo{pages}{036403} (\bibinfo{year}{2010}).

\end{thebibliography}
\end{document}